\documentclass[twocolumn]{article}

\usepackage{arxiv}
\usepackage[utf8]{inputenc} 
\usepackage[T1]{fontenc}    
\usepackage{hyperref}       
\usepackage{url}            
\usepackage{booktabs}       
\usepackage{amsfonts}       
\usepackage{nicefrac}       
\usepackage{microtype}      
\usepackage{graphicx}
\usepackage{natbib}
\usepackage{efbox}
\usepackage{amsmath}
\usepackage{multirow}
\usepackage{multicol}
\usepackage{makecell}
\usepackage{floatrow}

\title{TransMode-LLM: Feature-Informed Natural Language Modeling with Domain-Enhanced Prompting for Travel Behavior Modeling}

\author{
  Meijing Zhang \\
  Engineering Systems and Design\\
  Singapore University of Technology and Design\\
  8 Somapah Road, Singapore \\
  \texttt{meijing\_zhang@sutd.edu.sg} \\
   \And
  Ying Xu* \\
  Engineering Systems and Design\\
  Singapore University of Technology and Design\\
  8 Somapah Road, Singapore \\
  \texttt{xu\_ying@sutd.edu.sg} \\
}

\begin{document}
\twocolumn[
\maketitle
\begin{abstract}
Understanding traveler behavior and accurately predicting travel mode choice are at the heart of transportation planning and policy-making. Traditional statistical and machine learning methods struggle to capture the complex contextual factors and interdependencies underlying individual decision-making processes. Large language models (LLMs) provide new pathways to address these gaps by leveraging their pre-trained knowledge about human decision-making. This study proposes TransMode-LLM, an innovative framework that integrates statistical methods with LLM-based techniques to predict travel modes from travel survey data. The framework operates through three phases: (1) statistical analysis identifies key behavioral features, (2) natural language encoding transforms structured data into contextual descriptions, and (3) LLM adaptation predicts travel mode through multiple learning paradigms including zero-shot and one/few-shot learning and domain-enhanced prompting. We evaluate TransMode-LLM using both general-purpose models (GPT-4o, GPT-4o-mini) and reasoning-focused models (o3-mini, o4-mini) with varying sample sizes on real-world travel survey data. Extensive experiment results demonstrate that the LLM-based approach achieves competitive accuracy compared to state-of-the-art baseline classifiers models. Moreover, few-shot learning significantly improves prediction accuracy, with models like o3-mini showing consistent improvements of up to 42.9\% with 5 provided examples. However, domain-enhanced prompting shows divergent effects across LLM architectures. In detail, it is helpful to improve performance for general-purpose models with GPT-4o achieving improvements of 2.27\% to 12.50\%. However, for reasoning-oriented models (o3-mini, o4-mini), domain knowledge enhancement does not universally improve performance. This study advances the application of LLMs in travel behavior modeling, providing promising and valuable insights for both academic research and transportation policy-making in the future.
\end{abstract}
]

\keywords{Large Language Models; Travel Behavior Analysis; Mode Choice Prediction; Transportation Planning}

\section{Introduction}\label{sec1}

Understanding traveler behavior and accurately predicting individual travel mode choices are important for transportation planning and policy-making. The majority of previous studies on travel mode prediction are based on traditional statistical models like multinomial logit models to advanced machine learning techniques, both relying on structured data to establish relationships between traveler characteristics and mode choices \citep{zhao2020prediction,kashifi2022predicting}. While these traditional methods may capture statistical patterns, they struggle to capture the complex contextual factors and interdependencies underlying individual decision-making processes. More critically, these methods cannot reason about new or evolving situations, limiting their adaptability to changing circumstances or policy interventions. As a result, these traditional methods may yield suboptimal prediction accuracy and have difficulty in modeling how travelers adapt their choices in response to changing circumstances, infrastructure improvements, or policy interventions. The recent rapid development of Large Language Models (LLMs) provides a revolutionary opportunity to address these challenges. Trained on vast amounts of text data, LLMs demonstrate significant advantages in possessing remarkable capabilities in understanding context, reasoning about complex relationships, and generating human-like responses. 

Recently, researchers have shown an increased interest in applying the advanced techniques from LLMs to understand and predict human behavior across various domains including law, economics, political science, and social science \citep{shaghaghian2020customizing,aher2023using,argyle2023out,ziems2024can}. This trend is largely driven by the fact that these fields focus on language processing tasks with naturally textual data. In contrast, transportation research primarily represents behavioral information using categorical and numerical variables instead of natural language, which has limited the direct application of LLMs in this domain. Despite this gap, recent work in the transportation domain has begun to demonstrate the promising potential of LLM-based approaches for travel behavior prediction \citep{mo2023large,liu2024can,xu2025evaluating}. Nevertheless, a systematic investigation into adapting LLMs to travel mode prediction tasks remains absent. Drawing inspiration from this foundation, we propose TransMode-LLM, an innovative framework designed to predict travel modes by transforming structured behavioral data into natural language descriptions of travelers and their trips. Our methodology begins with statistical analysis that identifies key behavioral features through literature-based variable selection and feature importance analysis. We then transform these selected structured variables into narrative descriptions to enable LLMs to process and reason with contextual information that might be missed in traditional modeling approaches. To enhance the performance of LLMs for transportation-specific tasks, we explore various learning paradigms (zero-shot and one/few-shot learning) to understand their impact on travel mode prediction using natural language. We further propose a domain-enhanced prompting strategy that incorporates standardized mode definitions and three-step structured decision process. To validate the proposed LLM-based approach, we conduct experiments on the real-world dataset, in which we evaluate the performance of our model under various large language models across different sample sizes. Performance is assessed through three systematic comparisons: LLM-based approach versus traditional statistical classifiers, zero-shot versus few-shot learning paradigms, and performance with domain knowledge enhancement versus without it. The results demonstrate that the proposed LLM-based approach has the potential to yield competitive performance against established baseline models. The results also demonstrate that few-shot learning significantly improves prediction accuracy, with models like o3-mini showing consistent improvements of up to 42.9\% with 5 provided examples. In addition, domain-enhanced prompting is also helpful to improve the performance, with GPT-4o achieving improvements of 2.27\% to 12.50\%. To provide a comprehensive evaluation of the proposed method, we further build a hierarchical evaluation system complemented by F1 scores (F1-Macro and F1-Weighted) as additional measures of performance. F1 score analysis demonstrates that few-shot learning effectively addresses class imbalance, narrowing the accuracy-F1-macro gap by up to 71\%. By combining traditional travel behavior prediction methods with the contextual intelligence of LLMs, the proposed methodology in this study offers a novel pathway to advance travel mode prediction beyond conventional limitations. 

\section{Related Work}\label{sec2}

\subsection{Travel mode prediction}

Recent years have witnessed a growing academic interest in travel mode prediction. A number of techniques have been developed on mode prediction, such as traditional statistical models and advanced machine learning models. The majority of the studies in early years on travel mode prediction focused on discrete choice models such as multinomial logit and nested logit models. These models are based on random utility theory, which estimates the probability of a traveler choosing a particular mode based on the perceived utility among alternative modes. Utility is typically constructed as a function of various attributes, such as travel time, travel cost, and comfort \citep{ben1985discrete}. While these models provide theoretical foundation and interpretability, their assumptions on the independence of irrelevant alternatives and linear-in-parameters utility functions limit their ability to capture the complex interdependencies in real-world travel behavior \citep{zhao2020prediction}, which may not adequately capture the complexity of real-world travel behavior. To address the limitations of traditional discrete choice models, researchers have increasingly explored the application of advanced machine learning techniques in travel mode prediction. Evidence from a number of studies has confirmed the effectiveness of machine learning techniques in travel mode prediction \citep{hagenauer2017comparative,wang2018machine,ma2020travel,kashifi2022predicting}. However, both approaches remain limited because of their reliance on structured feature representations and these structured representations cannot capture the rich contextual complexity and interdependencies in travel decisions. Large language models can potentially address these gaps using their ability to process and reason with natural language descriptions of travel information.

\subsection{Large language models in travel mode prediction}

Large language models (LLMs), refer to transform-based language models containing hindered of billions or more of parameters trained on massive text data \citep{zhao2023survey}, such as GPT-3 \citep{brown2020language}, GPT-4 \cite{achiam2023gpt} and LLAMA \cite{touvron2023llama}. These models demonstrate strong natural language understanding capabilities and can solve complex tasks through text generation. The application of LLMs in the prediction of travel modes is an emerging research area with with initial promising results that has been identified \citep{mo2023large,liu2024can,xu2025evaluating}. 

\cite{mo2023large} first applied LLMs for the complex travel behavior prediction task without relying on traditional data-based training. The authors proposed a zero-shot prompting framework by incorporating task descriptions, travel characteristics, passenger attributes, and thinking guidance with domain knowledge. Using the Swissmetro stated preference dataset, LLM-based approach demonstrated the competitive predictive performance in accuracy and F1-score compared to traditional methods such as multinomial logit, random forest, and neural networks. However, \cite{liu2024can} pointed out that zero-shot performance still suffers from limited robustness. To address this limitation, they involved few-shot prediction and prediction based on persona loading in addition to zero-shot prediction method. Testing on the same Swissmetro stated preference dataset, they found that both methods improved the prediction performance on human behavior, with persona loading method yielding the best results among the three methods. \cite{xu2025evaluating} further studied the LLM-based travel mode choice prediction by introducing Retrieval-Augmented Generation (RAG). They evaluated four retrieval strategies: basic RAG, RAG with balanced retrieval, RAG with cross-encoder for re-ranking, and RAG with balanced retrieval and cross-encoder for re-ranking. They found that LLM-based approaches demonstrated superior generalization ability compared to traditional methods and the strategy combined with balanced retrieval and cross-encoder re-ranking achieved the highest prediction accuracy, significantly exceeding traditional baselines. These studies demonstrate the potential of LLMs in travel mode choice prediction. However, these efforts remain preliminary and no systematic framework currently exists for adapting LLMs to travel mode prediction tasks. This gap motivates us to systematically explore how LLMs can be effectively applied to structured transportation data through natural language transformation and domain-enhanced prompting strategies.

While these transportation-specific studies remain limited, broader applications of LLMs to behavioral understanding and prediction across different domains (i.e., economics, social science) demonstrate their capacity to simulate human decision-making \citep{aher2023using,ziems2024can,hegselmann2023tabllm,hussain2024tutorial}. \cite{aher2023using} introduced Turing Experiments to evaluate language models' ability to simulate different aspects of human behavior in experimental settings. The authors successfully replicated findings from classical behavioral studies (i.e., Ultimatum Game, Garden Path Sentences, and Milgram Shock Experiment). Their results demonstrated that larger models such as GPT-3 text-davinci-002 demonstrated the ability to faithfully reproduce complex human behavioral patterns including gender-sensitive decision-making and obedience to authority, thereby establishing LLMs as promising tools for behavioral analysis and prediction. \cite{ziems2024can} further comprehensively examined the potential of large language models in computational social science. By testing 13 language models on 25 representative English computational social science tasks including behavioral tasks, they found that LLMs achieved moderate-to-good agreement with humans on approximately half of classification tasks and demonstrated superior performance on generative explanation tasks. These findings suggest that LLMs can capture and predict human behavioral patterns, a premise that motivates our investigation into LLMs' applicability for travel mode choice prediction as a form of human decision-making behavior.

Given that data for travel mode choice are typically structured and tabular (e.g., household travel survey data), methodological advances that extend large language models (LLMs) to structured data are particularly relevant. \cite{hegselmann2023tabllm} introduced TabLLM framework, which transforms a serialization of tabular data into natural language representations for classification tasks. The authors demonstrated that LLMs can effectively leverage prior knowledge to predict complex behaviors across multiple domains (i.e., healthcare, finance, consumer, and gaming behavioral tasks) with minimal training data. \cite{hussain2024tutorial} focused on practical implementation and provided a systematic tutorial framework specifically designed for behavioral scientists, demonstrating how open-source LLMs can be effectively deployed to address research questions in behavioral science through appropriate experimental design and evaluation protocols. Inspired by these cross-domain successes, our work adapts these methodological insights to transportation, leveraging LLMs' demonstrated capacity for behavioral prediction.

\subsection{Research gaps and contributions}

Despite growing interest in applying LLMs to travel behavior prediction, several significant gaps remain in the current literature. First, previous work lacks a comprehensive feature selection analysis for natural language prompts \citep{mo2023large,liu2024can,xu2025evaluating}. As LLM-based approaches transforms structured tabular data into narrative descriptions, determining which features to include becomes important. This approach raises questions about prompt efficiency given token limits and potential noise from less relevant features. 

Second, systematic investigation of prompting strategies for travel mode prediction remains underdeveloped. While domain knowledge integration has been explored through general reasoning guidance \citep{mo2023large} and RAG strategies \citep{xu2025evaluating}, systematic investigation of how different prompting strategies transportation-specific knowledge such as standard mode definitions and structured decision frameworks can be integrated into prompts remains limited. 

Third, existing studies lack extensive comparative experiments and rigorous analysis across different contexts such as model configurations, dataset sizes, and evaluation metrics. Existing studies evaluate LLM performance on single dataset with limited comparative analysis \citep{mo2023large,liu2024can} or focus primarily on RAG strategy optimization within a single geographic region \citep{xu2025evaluating}. This will raise questions in generalizability and boundary conditions of LLM-based approaches, thereby limiting both theoretical understanding of the cases that LLM-based approaches are advantageous and practical guidance for their applications in transportation planning.

To address these gaps, we propose, TransMode-LLM, a novel framework designed to predict travel modes from natural language descriptions of travelers and their trips. Our methodology starts from travel survey data and then implements statistical analysis that identifies key behavioral features through literature-based variable selection and feature importance analysis. We then employ natural language encoding to transform these selected features into narrative descriptions to enable LLMs to process and reason with contextual information. Specifically, we implement multiple learning paradigms including zero-shot and one/few-shot learning, along with domain-enhanced prompting strategies, to enhance LLMs' prediction capabilities for travel mode prediction. To validate the proposed LLM-based approach, we conduct experiments on the real-world dateset. We evaluate performance through three systematic comparisons: LLM-based approaches versus traditional statistical classifiers, zero-shot versus one/few-shot learning paradigms, and performance with versus without domain knowledge enhancement. Performance is evaluated based on accuracy and F1 scores (F1-Macro and F1-Weighted) to provide comprehensive evaluation. 

\section{Problem Formulation} \label{sec3}

Travel mode choice prediction is traditionally formulated as a classification problem, where features describing personal, household, trip-specific, built environment, and economic factors are mapped directly to predict the transportation mode that a traveler will choose. However, this direct feature-to-mode mapping approach may not fully capture the complex contextual relationships and decision-making processes that humans naturally consider when choosing transportation modes. In this study, we propose a novel reformulation that leverages natural language processing to better capture these contextual relationships instead of directly mapping features to modes. In detail, the reformulated problem can be decomposed into two sequential steps: first, transforming structured features into coherent natural language descriptions, and second, applying LLMs for travel mode prediction based on these descriptions. 

Formally, the problem can be formulated as follows: 
Let $X = \{x_1, x_2, \ldots, x_n\}$ denote a set of features for a trip record, where each feature $x_i$ represents individual socio-demographic characteristics, household characteristics, trip characteristics, built environment factors, or pricing factors. These features are a mixture of numerical and categorical variables. We define a transformation function $D$ that converts this structured feature vector into a coherent natural language description:

\begin{equation}
D: X \rightarrow \mathcal{T}
\end{equation}
where $\mathcal{T}$ denotes the space of natural language descriptions relevant to travel mode choice. This transformation function is implemented as a template-based generator that follows a consistent narrative structure, ensuring that critical decision factors from the original features are preserved and contextualized in the linguistic representation.

Given this natural language representation, the prediction problem aims to estimate the conditional probability distribution $P(m|D(X))$ over transportation modes $m \in \mathcal{M} = \{\text{Car}, \text{Walk}, \text{Bus}, \ldots\}$. We define the function $\text{Prompt}(D(X))$ to construct the complete input by augmenting the natural language description $D(X)$ with essential task instructions and output formatting requirements, which is necessary to guide the model toward generating accurate and structured predictions. The probability distribution is obtained as:

\begin{equation}
P(m|D(X)) = \text{LLM}_{\alpha}(\text{Prompt}(D(X)), \theta)
\end{equation}
where $\alpha$ denotes the specific model architecture (e.g., GPT-4o, GPT-4o-mini, o3-mini, o4-mini), and $\theta$ represents the learning paradigm (e.g., zero-shot, one/few-shot). The predicted mode is then obtained as:
\begin{equation}
\hat{m} = \arg\max_{m \in \mathcal{M}} P(m|D(X))
\end{equation}
which selects the transportation mode with the highest predicted probability.

\section{Methods}\label{sec4}

Following Section \ref{sec3}, we now explain how our methodological framework transforms transportation data to enable LLM-based mode prediction. As illustrated in Figure \ref{fig:workflow}, the framework starts from travel survey data and operates through a systematic pipeline with three core phases. First, the original travel survey data go through feature analysis to identify which features most influence mode choice. Second, these selected features are transformed into contextual descriptions that LLMs can process. Lastly, with these contextual descriptions, general-domain LLMs are adapted through our prediction framework for the travel mode prediction task. The three core phases are formally designated as follows:

\begin{figure*}
  \centering
  \efbox{\includegraphics[width=\linewidth]{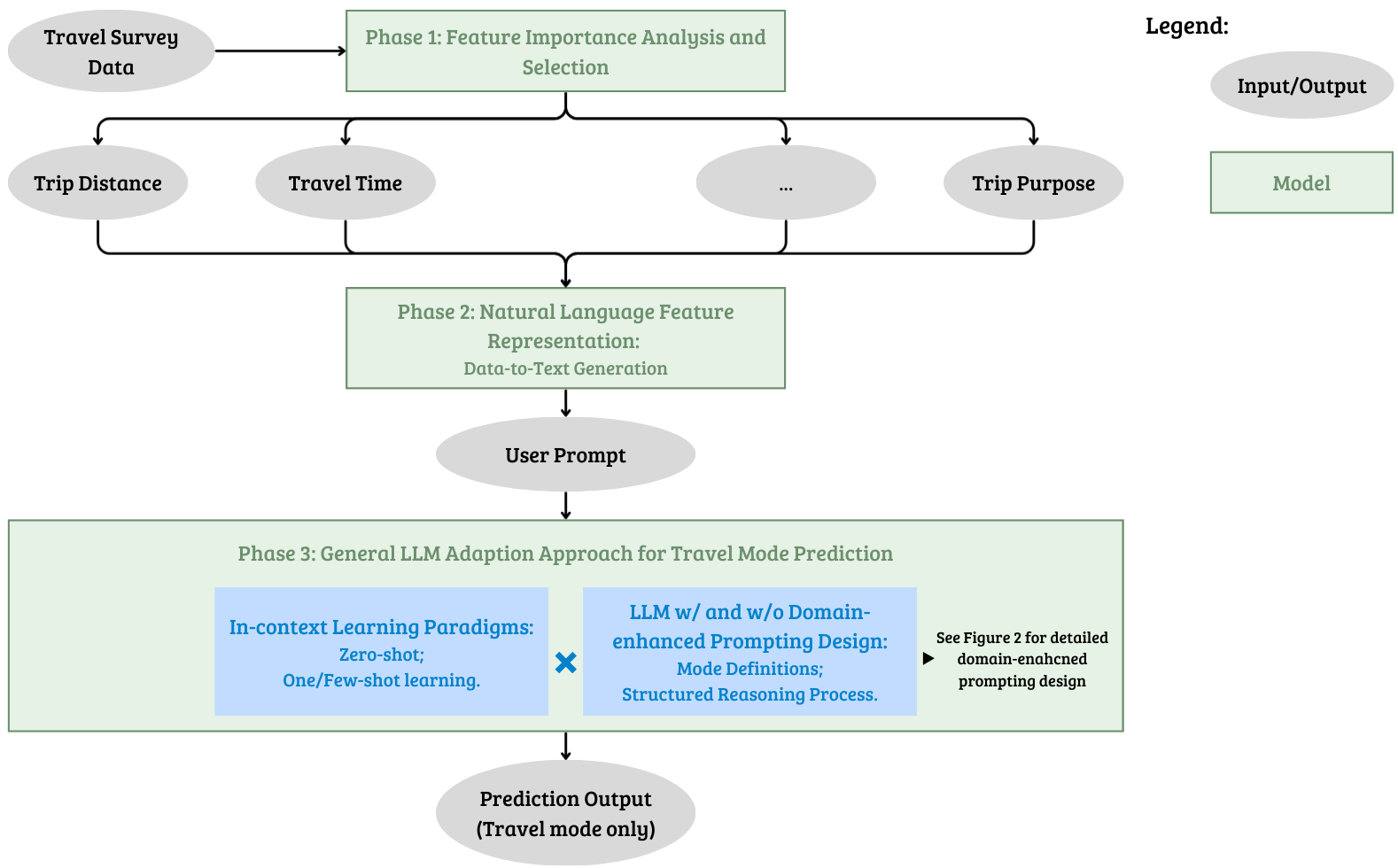}}
  \caption{Research Framework of LLM-Based Travel Mode Prediction: Our TransMode-LLM Model.}
  \label{fig:workflow}
\end{figure*}

\begin{itemize}
\item Phase 1 - Feature Importance Analysis and Selection: We apply multiple feature importance methods with ensemble ranking to identify the most predictive variables.
\item Phase 2 - Natural Language Feature Representation: We systematically transform the selected structured features into descriptive text representations that enable LLMs to perform contextual reasoning on travel mode prediction.
\item Phase 3 - General LLM Adaption Approach for Travel Mode Prediction: This phase adapts general LLMs for travel Mode prediction by implementing multiple learning paradigms (zero-shot and one-shot/few-shot) and domain enhanced prompting strategies domain-specific knowledge and structured reasoning processing methods, aiming to guide general LLMs to provide accurate travel mode predictions.
\end{itemize}

Next, we explain each phase in the following sections one by one. 

\subsection{Feature Importance Analysis and Selection}

Before entering the heart of the methodology by using natural language transformations to enable contextual reasoning by LLMs, in this section, we perform preliminary analysis to identify and prioritize impacting variables from the travel survey dataset. We first identify potentially impacting factors from the literature, followed by applying feature importance analysis to select the top 15 impacting factors. This two-stage selection process can help reduce the dimensionality to prevent overfitting and focuses computational resources, aiming to capture the most relevant factors of mode choice behavior.

\subsubsection{Impacting factors of travel behavior}

To perform travel mode choice prediction, a comprehensive understanding of the determinants influencing travel behavior is important. Zhou (2012) established a six-category framework for factors impacting mode choice: physical environment and urban form, mode-specific attributes, trip-makers' personal characteristics, trip parameters, travel demand management (TDM) measures, and psychological factors \cite{zhou2012sustainable}. Building upon this foundation and drawing from the methodological approach presented by Hörl and Balac (2021) \cite{horl2021synthetic}, this study explores five distinct categories of influential factors: individual socio-demographic characteristics, household characteristics, trip characteristics, built environment, and pricing factors. This refined categorization enables a more structured analysis of the complex interactions between personal, household, trip-specific, built environment, and economic variables that collectively shape travel behavior patterns.

\begin{itemize}
\item \textbf{Individual socio-demographic characteristics:}

Individual socio-demographic characteristics include mode use, distances traveled, travel frequency and etc. \cite{hanson1981travel,lu1999socio}. These characteristics pose a direct impact on travel behaviour and also impact travel behaviour indirectly via their impact on activity participation such as subsistence, maintenance, recreation and other.

\item \textbf{Household characteristics:}

A latest literature review conducted by Hu et al. (2023) indicates that there is a relationship between household information and individual activity and travel behaviors \cite{hu2023intra}. Other research also point out the similar findings \cite{hanson1981travel,lu1999socio,srinivasan2002travel}. The characteristics of household described in these studies include the number of children, number of workers, number of vehicles, household income, household type and location.

\item \textbf{Trip characteristics:}

To comprehensively capture the characteristics of the trips, three categories are included: geospatial context, temporal patterns and engaged activities during the trips. Geospatial context refers to the origin and destination of the trip, stops during the trips as well as the travel distances. Temporal patterns include the type of the travel day (weekday or weekend), departure and arrival times, stop time and travel duration. Regarding to the engaged activities, their relevant attributes typically include the overall trip purpose, the number of stops made during the journey and corresponding purposes.

\item \textbf{Built environment:}

Built environment thought to be influencing travel mode choice have been explored in several studies \cite{chen2008role,ewing2010travel,ding2017exploring,ko2019exploring}. Ewing and Cervero (2010), categorized the characteristics of the built environment that impacts travel mode choice as density, diversity, design, destination accessibility, distance to transit \cite{ewing2010travel}.

\item \textbf{Pricing factors:}

Economic considerations are an important contributory factor to the travel mode choice through direct financial incentives and constraints that shape travel mode decisions. The pricing factors embedded in transportation systems, particularly fuel prices and parking costs function as market signals that directly affect the relative affordability of different travel modes. Parking costs, considered as powerful travel demand management tools that can substantially alter mode choice decisions in urban environments where alternatives to private vehicle usage exist. Several studies have explored that the parking costs have an influence on travel mode choice \cite{hess2001effect,washbrook2006estimating,shoup2021high,vidovic2023impact}. Similarly, research have indicated the impact of fuel price on travel mode choice \cite{creutzig2014fuel,haire2007impact,jaggi2012modeling}.

\end{itemize}

\subsubsection{Feature importance analysis} \label{subsec:feature_importance}

To identify the most impacting factors on travel mode choice from the literature, we continue to conduct a systematic feature importance analysis. Following Cherepanova et al. (2023) which compares a number of feature-selection methods on tabular datasets \cite{cherepanova2023performance}, we apply multiple feature selection methods to overcome algorithm-specific biases and ensure robust identification of impacting factors. The analysis involves six basic methods (Univariate Statistical Test, Lasso, Random Forest, XGBoost, First-Layer Lasso, and Deep Lasso) and three advanced techniques (Adaptive Group Lasso, LassoNet, and FT-Transformer with Attention Map Importance).

The performance evaluation of our feature selection methods revealed that XGBoost achieved the highest classification accuracy (87.55\%), followed by Univariate Statistical Test (86.46\%) and First-Layer Lasso (86.01\%). By aggregating rankings across all algorithms, we identified the top 15 most impacting factors of travel mode choice. 

Table \ref{tab:all-factors} summarizes all candidate factors examined, their definitions, mean rankings, and final selection status. From this Table, it is obvious to see that trip distance emerges as the most consistently important impacting feature (mean ranking: 4.3), followed by travel time (8.3), age (8.3), and driving license status (8.7). In our ranking system, lower values indicate higher feature importance, as rankings represent average positions across all feature selection methods, with the most impacting factors receiving rankings closer to 1. This ranking-based aggregation approach provides robustness by synthesizing results from multiple feature selection algorithms, reducing dependence on any single method's biases.

\begin{table*}[htbp]
\centering
\caption{Summary of factors impacting travel behavior and feature selection results}
\label{tab:all-factors}
\begin{tabular}{p{0.14\textwidth} p{0.20\textwidth} p{0.40\textwidth} p{0.10\textwidth} p{0.08\textwidth}}
\toprule
\textbf{Category} & \textbf{Factor} & \textbf{Definition} & \textbf{Mean Ranking} & \textbf{Selected} \\
\midrule
\multirow{7}{*}{\makecell[l]{Trip \\characteristics}} 
& Travel distance & Trip distance in miles & 4.3 & $\checkmark$ \\
& Travel duration & Trip duration in minutes & 8.3 & $\checkmark$ \\
& Trip purpose & General purpose of trip (e.g., work, shopping, social) & 11.6 & $\checkmark$ \\
& Trip purpose (specific) & Detailed trip purpose category & 14.7 & $\times$ \\
& Day type & Whether trip occurred on weekend or weekday & 17.9 & $\times$ \\
& Start time & Trip start time in 24-hour format & 16.1 & $\times$ \\
& End time & Trip end time in 24-hour format & 15.3 & $\times$ \\
\midrule
\multirow{4}{*}{\makecell[l]{Person \\characteristics}}
& Age & Respondent age in years & 8.3 & $\checkmark$ \\
& Gender & Respondent sex & 9.3 & $\checkmark$ \\
& Driving license status & Whether respondent holds a valid driver's license & 8.7 & $\checkmark$ \\
& Employment status & Whether respondent is employed & 11.0 & $\checkmark$ \\
\midrule
\multirow{6}{*}{\makecell[l]{Household \\characteristics}}
& Household size & Total number of people in household & 10.3 & $\checkmark$ \\
& Number of vehicles & Total number of vehicles in household & 11.3 & $\checkmark$ \\
& Homeownership & Whether home is owned, rented, or other arrangement & 11.6 & $\checkmark$ \\
& Household income & Annual household income & 10.7 & $\checkmark$ \\
& Number of drivers & Number of licensed drivers in the household & 13.6 & $\times$ \\
& Number of workers & Number of employed persons in household & 14.4 & $\times$ \\
\midrule
\multirow{4}{*}{Built Environment} 
& Urban/Rural designation & Whether household is located in urban or rural area & 9.4 & $\checkmark$ \\
& Population size & MSA population size category from five-year ACS & 13.4 & $\checkmark$ \\
& Rail availability & Whether MSA has heavy rail transit access & 11.6 & $\checkmark$ \\
& Gasoline price & Weekly regional gasoline price in cents during travel day & 12.0 & $\checkmark$ \\
& Life cycle classification & Household life cycle classification & 16.0 & $\times$ \\
\midrule
\multirow{1}{*}{Pricing factors}
& Parking costs & Amount paid for parking & 16.3 & $\times$ \\
\bottomrule
\end{tabular}
\end{table*}

\subsection{Natural Language Feature Representation: Data-to-Text Generation}

To enable LLM processing, it is necessary to transform the selected tabular features into natural language descriptions. Following the feature selection process described in Section 4.1, we transform the 15 selected features into structured text descriptions for each trip record. This transformation converts each row (selected feature) of tabular data into a narrative description that presents individual and household characteristics, trip attributes, built environment and pricing factors as an integrated scenario. The narrative format allows LLMs to leverage their pre-trained semantic understanding of travel concepts and contextual relationships between features. Each trip is described using a fixed template as illustrated below, in which bold text indicating the 15 selected features.

\begin{quote}
\textit{Consider a \textbf{44-year-old female} who is a \textbf{driver} and is \textbf{employed}. She is living in a household with \textbf{3 people}, and \textbf{1 vehicle}, with a \textbf{household income} of \$125,000 to \$149,999, in a home that is \textbf{owned with a mortgage}. She is traveling for \textbf{shopping}, for a \textbf{distance} of 1.3 miles, with an expected \textbf{travel time} of 10 minutes. She lives in an \textbf{urban area}, with \textbf{no access to rail transit}, in an \textbf{MSA} of 500,000 to 999,999, where the \textbf{gas price} is \$4.30 per gallon. What is the most likely transportation mode she would choose for this trip?}
\end{quote}

The resulting descriptions provide a comprehensive narrative that captures the essential context of each trip while preserving the information content of the original feature vector. This approach allows the LLMs to process transportation data in a format that aligns with its pre-training, enabling more nuanced interpretation of the relationships between features.

\subsection{General LLM Adaption Approach for Travel Mode Prediction}

Building upon the natural language feature representations developed in Section 4.2, we now turn to the key challenges of applying general-domain LLMs to travel mode prediction. Despite their impressive language understanding capabilities, general-domain LLMs are not naturally optimized for travel mode prediction tasks due to the fact that LLMs lack exposure to domain-specific terminology, trip-mode compatibility constraints and the structured reasoning patterns required for accurate prediction. To address these limitations, we propose a comprehensive framework employing two domain specialization techniques (1) in-context learning paradigms and (2) domain-enhanced prompting design. In-context learning paradigms provide task demonstration examples within prompts, including zero-shot, one-shot, and few-shot configurations to enable LLM adaptations. Domain-enhanced prompting design provides standardized mode definitions and structured reasoning processes to explicitly inject transportation knowledge into the model. We provide more details about these two techniques in Section \ref{subsec:learning_paradigms} and Section \ref{subsec:prompting_design} respectively. In addition, in Section \ref{subsec:model_selection}, we introduce the rationale and criteria for selecting LLMs for systematic comparison.

\subsubsection{In-context Learning Paradigms}
\label{subsec:learning_paradigms}

In-context learning paradigms can address the domain adaptation challenges by providing task-specific demonstration examples that can compensate for the underrepresentation of travel mode prediction in general LLM training corpora. The demonstration examples are provided at inference time as conditioning. Following Brown et al. \cite{brown2020language}, we implement three learning paradigms based on demonstration example availability to investigate the impact of providing demonstration examples: zero-shot learning (no demonstration example), one-shot learning (single demonstration example), and few-shot learning (multiple demonstration examples).

Zero-shot learning tests the LLMs' inherent capabilities by making predictions without any task-specific training or examples, relying solely on pre-trained knowledge and contextual understanding of natural language descriptions.  In contrast, one-shot and few-shot learning involve providing a single and a small number of task-specific demonstrations at inference time as conditioning without permitting weight updates, respectively \cite{brown2020language}. We implement one-shot learning with a single example and few-shot learning with 2, 3, 5, and 10 examples to systematically evaluate how demonstration quantity impact prediction performance. For one-shot and few-shot configurations, we select demonstration examples from the training data using stratified sampling to ensure travel mode representation. The selection strategy prioritizes mode diversity by first sampling at least one example per travel mode, then balancing additional examples proportionally to mode frequency in the training data, ensuring diverse trip characteristics while maintaining representativeness. 

\subsubsection{Domain-enhanced Prompting Design}
\label{subsec:prompting_design}

\begin{figure}
  \centering
  \includegraphics[width=\linewidth]{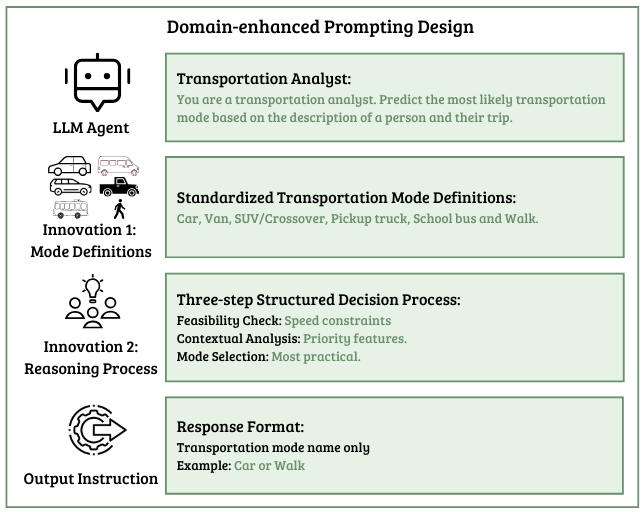}
  \caption{Domain-enhanced Prompting Design of LLM-Based Travel Mode Prediction.}
  \label{fig:prompt_design}
\end{figure}

While in-context learning paradigms can improve domain adaptation through task-specific demonstrations, general LLMs still face fundamental limitations in transportation applications. They may lack travel behavior terminology and concepts due to insufficient transportation-related knowledge in their pretraining data. Furthermore, general LLMs lack structured reasoning frameworks for systematically evaluating travel decisions. To address these challenges, we develop a domain-enhanced prompting framework that positions LLMs as virtual transportation analysts capable of predicting travel modes from natural language trip descriptions. Our framework incorporates two key technical innovations: (1) standardized transportation mode definitions that provide consistent domain knowledge, and (2) a three-step structured decision process that guides systematic reasoning. Figure \ref{fig:prompt_design} illustrates the complete architecture of this prompting design. The following subsections detail each innovation. 

The first innovation is to incorporate standardized mode definitions inspired by the National Household Travel Survey \citep{bricka2024summary}. Pre-testing through Chatbox AI \citep{chatboxai}, revealed that LLMs without domain knowledge misinterpret 'car' as a broad vehicle category, failing to distinguish between vans, SUVs/crossovers, and pickup trucks.  By introducing standardized definitions (shown in the "Mode Definitions" panel of Figure \ref{fig:prompt_design}), we can provide explicit transportation terminology that enables the model to correctly differentiate vehicle types and other mode categories.

The second innovation is to develop a machine learning-informed three-step chain-of-thought reasoning process that structures predictions according to empirically-derived feature hierarchies. As shown in Figure \ref{fig:prompt_design}, the reasoning process first starts with Feasibility Check (Step 1) which computes travel speed required (distance/time) and remove physically infeasible modes based on speed constraints. In Step 2, Contextual Analysis guides the model to evaluate factors in their empirically-determined order of importance into the chain-of-thought structure in accordance with the feature importance rankings from the systematic feature selection analysis in Section \ref{subsec:feature_importance}. For instance, trip distance (mean rank 4.3) and travel time (mean rank 8.3) emerged as the most influential features for predicting travel mode choice. Finally, Mode Selection (Step 3),directs the model to integrate all evaluated factors and select the most practical transportation mode, ensuring that domain constraints are respected (e.g., school bus only for school-related trips). Wei et al. (2022) demonstrate that chain-of-thought prompting is more effective than others for challenging and multi-step text-to-text tasks \citep{wei2022chain}. Given that travel mode prediction requires integrating multiple information dimensions, the proposed three-step framework extends this concept by incorporating domain-specific feasibility constraints and empirically-derived feature priorities, providing the structured reasoning pathway necessary for accurate travel mode prediction.

\subsubsection{Model Selection}
\label{subsec:model_selection}

The rapid development of large language models has produced a diverse landscape of models varying in scale, architectural design, and functional specialization. General-purpose models such as GPT-4o \citep{openai2024gpt4o} are designed to handle a wide range of tasks, while their compact variants like GPT-4o-mini offer resource-efficient inference with reduced computational costs \citep{openai2024gpt4omini}. More recently, reasoning-specialized architectures have emerged, including the O-series models (e.g., o3-mini and o4-mini), which represent third and fourth generation reasoning frameworks explicitly optimized for multi-step logical inference \citep{openai2024gpto3mini,openai2024gpto4mini}. In this study, we selected four models across these two categories: (1) GPT-4o and GPT-4o-mini as general-purpose models of different scales, and (2) o3-mini and o4-mini as reasoning-specialized compact models. With the diverse model selections, we aim to address the following research questions: Does model size within general-purpose architectures correlate directly with predictive performance? Do reasoning-specialized models offer advantages in capturing the complex decision-making processes underlying travel behavior, irrespective of their parameter count? Furthermore, what are the cost-effectiveness trade-offs between comprehensive and compact models for applications in travel mode prediction tasks? Through systematic comparisons across these models, we assess the relationship between model size and prediction accuracy, evaluate whether reasoning-focused architectures better capture travel behavior complexities, and identify cost-effective solutions for practical deployment in transportation contexts. 

\section{Experiments}

\subsection{Data preparation}

The dataset used in this study is the National Household Travel Survey conducted by the Federal Highway Administration (FHWA), which is a periodic national survey data supporting transportation planners and policymakers in their decision-making \cite{fhwa2022nextgen}. The performance of travel behavior analysis relies on the quality of input data, particularly when developing generative and predictive models for travel mode choice. As such, we first implement a comprehensive two-stage filtering methodology to address inconsistencies within the National Household Travel Survey (NHTS) dataset that could otherwise compromise model validity. 

The first step was applied a speed-distance-time consistency filter to identify physically implausible trip records such as walking trips with calculated speeds exceeding 5 mph or vehicle trips with unrealistically low speeds. Subsequently, a socio-demographic consistency filter targeted logical inconsistencies between the characteristics of the individuals and the reported travel behaviors, such as underage drivers, non-drivers operating vehicles, young children traveling unaccompanied on public transit, and private vehicle trips without an identified driver. After filtering, the final dataset contains 22,868 trips with 85 attributes. Statistical validation demonstrates that the filtered dataset exhibits more realistic distributions across all travel modes, with particularly significant improvements in walking, cycling, and public transit categories. By addressing both physical and logical inconsistencies, these preprocessing steps significantly enhances data reliability for travel behavior analysis, enabling more accurate prediction of travel mode choice.

\subsection{Experimental Design and Implementation} \label{sub:exp_design}

We evaluate the performance of LLM-based approach for travel mode prediction tasks across different learning paradigms and prompting strategies. The four models described in Section~\ref{subsec:model_selection} are examined across four sample sizes (100, 200, 500, and 1,000). To ensure practical relevance and computational efficiency, we focus on six (out of 20) travel modes: Car, Van, SUV/Crossover, Pickup truck, School bus, and Walk in this study which account for over 90\% of recorded trips and coverage typical travel patterns. As baselines, we select the high-performing machine learning classifiers LogitBoost and Gradient Boosting identified by \citep{wang2020predicting} as consistently superior performers among 86 classifiers from 14 model families evaluated on the National Household Travel Survey (NHTS) 2017 dataset, which is structurally analogous to the used dataset in this study. Moreover, to guarantee fair comparison between the two baseline methods (LogitBoost and Gradient Boosting) and LLM-based approaches, we use identical test sets across all methodologies within each sample size configuration. 

Regarding the details of LLMs' implementation, the temperature is set to 0 for models that support deterministic outputs (GPT-4o, GPT-4o-mini) while the o-series models (o3-mini and o4-mini) do not support temperature adjustment as their reasoning architecture uses fixed sampling parameters. For one/few-shot learning, we examine its effectiveness using 1, 2, 3, 5 and 10 demonstration examples as described in Section~\ref{subsec:learning_paradigms}. The selection of few-shot examples follows a stratified sampling strategy designed to maximize representational diversity and mode coverage. The experiments are all implemented using Python and the OpenAI API.

\subsection{Model Evaluation}

Travel mode modeling and prediction tasks are considered as classification problems \cite{hagenauer2017comparative}. Following this, we evaluate model performance using three standard classification metrics: accuracy, $\text{F1}_{\text{macro}}$ and $\text{F1}_{\text{weighted}}$ \citep{rainio2024evaluation}.

\subsubsection{Evaluation metric: accuracy}

Accuracy measures the overall percentage of correctly predicted travel modes. This metric provides a standard benchmark in classification tasks because it captures the overall percentage of correct predictions and provides a concise measure of aggregate model performance. 

Formally, let $\hat{y}_i$ denote the predicted travel mode for sample $i$ while $y_i$ denotes the actual mode. We let $\mathcal{D}^{\text{test}}$ represent the test dataset, and $M$ denote the set of all travel modes. Then the accuracy metric is formally defined as:

\begin{align}
    \text{Accuracy} &= \frac{1}{|\mathcal{D}^{\text{test}}|} \sum_{i \in \mathcal{D}^{\text{test}}} \mathbf{1}[\hat{y}_i = y_i] \label{eq:accuracy}\\
\end{align}

where $\mathbf{1}[\cdot]$ is the indicator function.

However, accuracy performs poorly when classes are imbalanced. The dataset used in this study is highly imbalanced, with more than one third of trips by car (37.9\%), followed by SUVs/crossovers (33.5\%), pickup trucks (10.7\%), and vans (6.0\%), accounting together for nearly 88\% of all observations. In contrast, other modes such as walking (6.8\%) and school bus (2.1\%) each represent only a small fraction of the data. This imbalance is common in transportation studies, where certain modes dominate travel behavior. To address this imbalance, we employ F1-macro and F1-weighted scores together to capture model performance across both minority and majority modes. Unlike overall accuracy, which can be misleadingly high by simply predicting the dominant modes, the combined use of F1-macro and F1-weighted scores can comprehensively evaluate model performance across all travel modes.

\subsubsection{F1-macro and F1-weighted scores}

F1-macro provides equal weight to each mode, making it particularly valuable for evaluating underrepresented but policy-relevant modes (e.g. walking, cycling). In contrast, F1-weighted weights performance by mode prevalence, thus reflecting overall population-level prediction accuracy.

Let $M$ denote the set of travel modes and $\text{F1}_m$ the F1-score for mode $m$. The overall F1-macro and F1-weighted scores are computed as:

\begin{align}
    \text{F1}_{\text{macro}} &= \frac{1}{|M|} \sum_{m \in M} \text{F1}_m \label{eq:f1_macro}\\
    \text{F1}_{\text{weighted}} &= \sum_{m \in M} \frac{|M_m|}{|\mathcal{D}^{\text{test}}|} \text{F1}_m \label{eq:f1_weighted}
\end{align}

where $M_m$ represents the subset of test samples with travel mode $m$.

Each $\text{F1}_m$ is derived from mode-specific precision and recall:

\begin{align}
    \text{F1}_m &= \frac{2 \cdot \text{Precision}_m \cdot \text{Recall}_m}{\text{Precision}_m + \text{Recall}_m} \label{eq:f1_score}\\
    \text{Precision}_m &= \frac{\text{TP}_m}{\text{TP}_m + \text{FP}_m} \label{eq:precision}\\
    \text{Recall}_m &= \frac{\text{TP}_m}{\text{TP}_m + \text{FN}_m} \label{eq:recall}
\end{align}
where $\text{TP}_m$, $\text{FP}_m$, and $\text{FN}_m$ denote true positives, false positives, and false negatives for travel mode $m$, respectively.

Precision captures prediction accuracy (correct predictions among all predicted), while recall measures completeness (correctly identified among all actual instances). For instance, 80\% precision for walking means that 80\% of trips predicted as walking are correct, with 20\% being false positives.

Given the imbalance in our dataset, a precision–recall trade-off arises:
\begin{itemize}
    \item \textbf{High precision, low recall:} Conservative predictions reduce false positives but miss many true cases, underestimating demand.
    \item \textbf{Low precision, high recall:} Liberal predictions capture most true instances but overestimate demand through false positives.
\end{itemize}

In summary, F1-macro and F1-weighted scores provide complementary perspectives: F1-macro reveals whether models perform equitably across all modes while F1-weighted reflects overall predictive accuracy aligned with observed mode distributions. Using both metrics in the following analysis is helpful to evaluate whether accuracy improvements translate into balanced performance or primarily benefit dominant travel modes.

\section{Results}

This section evaluates LLM-based approaches for travel mode prediction using two performance indicators: accuracy and F1 scores (F1-macro and F1-weighted). For each indicator, we present the prediction performance of LLM-based approaches through three key comparisons: (1) LLM-based approaches versus baseline classifiers (LogitBoost and Gradient Boosting), (2) zero-shot versus one/few-shot learning paradigms, and (3) performance with domain knowledge enhancement versus without it. Each comparison considers four language models (GPT-4o, GPT-4o-mini, o3-mini, o4-mini) across four sample sizes (100, 200, 500, and 1000). Section \ref{sub:accuracy} reports results of accuracy indicator, followed by F1 scores indicator in Section \ref{sub:f1scores}.

\subsection{Accuracy} \label{sub:accuracy}

The primary evaluation metric in this study is accuracy, the overall proportion of correct predictions compared to the actual travel mode from the dataset. The following sections present accuracy results through three comparisons: LLM-based approaches versus baseline classifiers, zero-shot versus one/few-shot learning paradigms, and performance with domain knowledge enhancement versus without it.

\subsubsection{Performance Comparison: LLMs vs. Baseline Classifiers}

To investigate whether LLM-based approaches are competitive for travel mode prediction tasks, we compare zero-shot LLM performance against baseline classifiers (LogitBoost and Gradient Boosting) as described in Section~\ref{sub:exp_design}. The results of the predictive accuracy performance are summarized in Table \ref{tab:baseline_vs_zeroshot}.

These results establish that LLM-based approaches, particularly advanced reasoning models such as o4-mini, achieve competitive or superior performance against established baseline classifiers without requiring any task-specific training. This represents a fundamental paradigm shift in predictive modeling for transportation applications, which redirects from training-based to LLM prompt-based prediction. 

While traditional classifiers may require extensive model training, feature engineering pipelines, hyperparameter optimization as well as ongoing maintenance infrastructure, LLMs achieve comparable or superior accuracy for travel mode choice prediction tasks through zero-shot inference alone using only natural language descriptions of the trips. Remarkably, o4-mini achieves this despite being trained on general web-scale data without transportation-specific knowledge, demonstrating that large-scale pre-training enables transferable reasoning capabilities that generalize to specialized domains.

\begin{table*}[htbp]
\centering
\caption{Comparison of baseline classifiers and LLM-based zero-shot performance}
\label{tab:baseline_vs_zeroshot}
\begin{tabular}{llcccc}
\toprule
\multirow{2}{*}{\textbf{Method}} & \multirow{2}{*}{\textbf{Model}} & \multicolumn{4}{c}{\textbf{Sample Size}} \\
\cmidrule(lr){3-6}
& & \textbf{100} & \textbf{200} & \textbf{500} & \textbf{1000} \\
\midrule
\multirow{2}{*}{\textbf{Baseline Classifiers}} 
& LogitBoost & \textbf{0.5100$\pm$0.058} & 0.4400$\pm$0.030 & 0.4580$\pm$0.034 & 0.4810$\pm$0.022 \\
& Gradient Boosting & 0.3700$\pm$0.060 & 0.4050$\pm$0.058 & 0.5040$\pm$0.035 & 0.5040$\pm$0.049 \\
\midrule
\multirow{4}{*}{\textbf{LLM-Based (Zero-Shot)}}
& GPT-4o & 0.4000 & 0.4750 & 0.4400 & 0.4500 \\
& GPT-4o-mini & 0.3500 & 0.4750 & 0.4000 & 0.4350 \\
& o3-mini & 0.3500 & 0.4500 & 0.4400 & 0.4500 \\
& o4-mini & 0.4500 & \textbf{0.5000} & \textbf{0.5600} & \textbf{0.5750} \\
\bottomrule
\end{tabular}
\end{table*}

\subsubsection{Performance Comparison: Zero-shot vs One/Few-shot Learning Paradigms} \label{sub:accuracy_zero_few}

With the establishment that LLMs under zero-shot conditions outperform baseline classifiers, we now turn to examining the impact of few-shot learning on travel mode choice prediction performance. Because LLMs are trained on broad datasets, they may still struggle with domain-specific tasks like travel mode prediction, as their attention mechanisms are optimized for general rather than specialized knowledge. Drawing from \cite{brown2020language}, few-shot learning provides in-context examples that guide LLM predictions through demonstration rather than parameter updates, potentially compensating for limited domain-specific knowledge in general-purpose models. 

Table \ref{tab:accuracy_few_shot_best} presents a systematic evaluation of few-shot learning efficacy across multiple large language models and sample sizes. For each configuration, we identify the optimal number of examples (ranging from 1 to 10) and compare this best-performing few-shot setup against the corresponding zero-shot baseline. Obviously, few-shot learning demonstrates considerable improvements in prediction accuracy compared to zero-shot baselines in most cases. The approach yields improvements in 13 of 16 experimental configurations, with performance gains ranging from 3.6\% to a remarkable 42.9\%. Notably, optimal performance is typically achieved with a modest number of examples (1-5), suggesting that even minimal contextual guidance can substantially enhance model performance. However, we also find that few-shot learning does not universally outperform zero-shot approaches as expected. A particularly instructive case occurs under the o4-mini model at a sample size of 1,000, where the inclusion of a single example reduces accuracy from 57.50\% to 54.00\%, a 6.1\% performance decrease. This suggests that highly capable reasoning models may already have the ability to possess sufficient domain knowledge to achieve best performance even without additional examples provided. Excessive examples may actually constrain the model's reasoning pathways in ways that diminish its inherent capabilities.

This pattern also aligns with findings in prior literature \citep{brown2020language,zhao2021calibrate}. \cite{zhao2021calibrate} pointed out that seemingly minor methodological decisions such as prompt formatting, example selection, and ordering can pose a significant impact on few-shot learning performance, with accuracy ranging from near-chance levels to state-of-the-art performance. This variability suggests that systematic prompt engineering strategies, particularly adaptive example selection based on model capability, may have the potential to improve few-shot learning reliability and may warrant dedicated investigation, which can be further investigated in future research.

\begin{table*}[htbp]
\centering
\caption{Best few-shot performance vs zero-shot baseline without domain knowledge enhancement}
\label{tab:accuracy_few_shot_best}
\begin{tabular}{lcccccc}
\toprule
\multirow{2}{*}{\textbf{Model}} & \multirow{2}{*}{\textbf{Sample Size}} & \textbf{Zero-Shot} & \textbf{Best Few-Shot} & \textbf{No.of} & \textbf{Improvement} \\
& & \textbf{Accuracy} & \textbf{Accuracy} & \textbf{Examples} & \textbf{(\%)} \\
\midrule
\multirow{4}{*}{GPT-4o}
& 100 & 0.4000 & 0.4000 & 1 & 0.0 \\
& 200 & 0.4750 & \textbf{0.5250} & \textbf{1} & \textbf{+10.5} \\
& 500 & 0.4400 & 0.4400 & 1 & 0.0  \\
& 1000 & 0.4500 & \textbf{0.4700} & \textbf{2} & \textbf{+4.4}  \\
\midrule
\multirow{4}{*}{GPT-4o-mini}
& 100 & 0.3500 & \textbf{0.4500} & \textbf{2/3} & \textbf{+28.6 } \\
& 200 & 0.4750 & 0.4750 & 2 & 0.0  \\
& 500 & 0.4000 & \textbf{0.4300} & \textbf{5} & \textbf{+7.5}  \\
& 1000 & 0.4350 & \textbf{0.4550} & \textbf{2/3/5} & \textbf{+4.6}  \\
\midrule
\multirow{4}{*}{o3-mini}
& 100 & 0.3500 & \textbf{0.5000} & \textbf{5} & \textbf{+42.9 } \\
& 200 & 0.4500 & \textbf{0.4750} & \textbf{1/3/5/10} & \textbf{+5.6}  \\
& 500 & 0.4400 & \textbf{0.4800} & \textbf{2} & \textbf{+9.1}  \\
& 1000 & 0.4500 & \textbf{0.4900} & \textbf{5} & \textbf{+8.9}  \\
\midrule
\multirow{4}{*}{o4-mini}
& 100 & 0.4500 & 0.5000 & \textbf{5} & \textbf{+11.1}  \\
& 200 & 0.5000 & 0.5500 & \textbf{3} & \textbf{+10.0}  \\
& 500 & 0.5600 & 0.5800 & \textbf{5} & \textbf{+3.6}  \\
& 1000 & \textbf{0.5750} & 0.5400 & 1 & -6.1 \\
\bottomrule
\end{tabular}
\end{table*}

\subsubsection{Performance Comparison: LLMs with Domain-knowledge Enhancement versus without it} \label{sub:accuracy_with_witho_domain}

Having investigated the few-shot learning effects on LLM-based travel mode prediction performance, we now move on to explore the impact of domain knowledge integration on the prediction performance. As introduced in Section 4.3, we propose a domain-enhanced prompting strategy that incorporates standard transportation mode definitions and the ranking of factors in order of importance into the prompting design. Table \ref{tab:few_shot_baseline_vs_enhanced} presents comparative results across four large language models under varying sample sizes, contrasting standard few-shot learning with domain-enhanced approaches.

The results show a striking architectural divergence in domain knowledge utilization. From Table \ref{tab:few_shot_baseline_vs_enhanced}, we can see that domain knowledge enhancement produces predominantly positive or neutral responses in general-purpose models such as GPT-4o and GPT-4o-mini, while causing performance degradation in reasoning-series models such as o3-mini and o4-mini. In detail, general-purpose models demonstrate a 75\% success rate (6 out of 8 cases showing positive or neutral effects), with GPT-4o successfully exemplifying the domain knowledge integration by achieving consistent improvements across all sample sizes (+2.27\% to +12.50\%). This universal benefit suggests that GPT-4o's architecture, which was designed for broad applicability across diverse tasks, has inherent mechanisms for harmonizing multiple knowledge sources. Conversely, the reasoning-series models such as o3-mini and o4-mini present a contrasting pattern, exhibiting only a 25\% success rate, with negative effects of providing domain knowledge observed in 5 out of 8 cases. The largest decrease in terms of prediction occurs with o3-mini at a sample size of 100, -20.00\% when domain knowledge is introduced. This suggests that the reasoning-series architectures might be exploiting specialized computational pathways that get disrupted by any external domain information.

The negative transfer observed in these reasoning models can be interpreted through the lens of architectural specialization. Models optimized for logical reasoning and problem-solving likely develop highly structured internal representations and inference mechanisms. The introduction of domain knowledge may interfere with these carefully tuned reasoning pathways, creating conflicting signals that degrade performance rather than enhance it. This interpretation is consistent with findings pointed out by \cite{wang-etal-2025-decoupling}, which shows that injecting external knowledge into models with strong reasoning capabilities can cause them to deviate from their original chain-of-thought reasoning paths, resulting in performance degradation. Similarly, \cite{amiraz2025distracting} demonstrate that even semantically relevant information can function as hard distractors that paradoxically impair LLM reasoning rather than enhance it. For example, o4-mini, optimized for efficient reasoning, may already have the ability of extracting relevant patterns from data without additional contextual guidance. The added domain knowledge may introduce noise or misalignment to the model's internal representations thereby posing a negative impact on the performance i.e. predictive accuracy. Thus, while domain knowledge enhancement serves as a valuable augmentation for general-purpose models (i.e. GPT-4o, GPT-4o-mini) that benefit from explicit guidance, it becomes a source of interference for reasoning-specialized architectures that have evolved sophisticated internal mechanisms for autonomous pattern discovery and logical inference.

\begin{table*}[htbp]
\centering
\caption{Few-shot performance with and without domain knowledge enhancement}
\label{tab:few_shot_baseline_vs_enhanced}
\begin{tabular}{lcccc}
\toprule
\multirow{2}{*}{\textbf{Model}} & \multirow{2}{*}{\textbf{Sample Size}} & \textbf{Without Domain} & \textbf{With Domain} & \textbf{Improvement} \\
& & \textbf{Best Few-Shot (Examples)} & \textbf{Best Few-Shot (Examples)} & \textbf{(\%)} \\
\midrule
\multirow{4}{*}{GPT-4o}
& 100 & 0.4000 (1) & \textbf{0.4500 (10)} & \textbf{+12.50} \\
& 200 & 0.5250 (1) & \textbf{0.5500 (5)}  & \textbf{+4.76} \\
& 500 & 0.4400 (1) & \textbf{0.4500 (2)}  & \textbf{+2.27} \\
& 1000 & 0.4700 (2) & \textbf{0.4850 (3)} & \textbf{+3.19} \\
\midrule
\multirow{4}{*}{GPT-4o-mini}
& 100 & \textbf{0.4500 (2/3)} & 0.4000 (1) & -11.11 \\
& 200 & 0.4750 (2) & 0.4750 (5) & 0.00 \\
& 500 & 0.4300 (5) & \textbf{0.4900 (5)} & \textbf{+13.95} \\
& 1000 & 0.4550 (2/3/5) & \textbf{0.4600 (1)} & \textbf{+1.10} \\
\midrule
\multirow{4}{*}{o3-mini}
& 100 & \textbf{0.5000 (5)} & 0.4000 (1) & -20.00 \\
& 200 & 0.4750 (1/3/5/10) & \textbf{0.5000 (3)} & \textbf{+5.26} \\
& 500 & 0.4800 (2) & 0.4800 (3) & 0.00 \\
& 1000 & \textbf{0.4900 (5)} & 0.4750 (1) & -3.06 \\
\midrule
\multirow{4}{*}{o4-mini}
& 100 & \textbf{0.5000 (5)} & 0.4500 (1) & -10.00 \\
& 200 & 0.5500 (3) & \textbf{0.5750 (3)} & \textbf{+4.55} \\
& 500 & \textbf{0.5800 (5)} & 0.5500 (2) & -5.17 \\
& 1000 & \textbf{0.5400 (1)} & 0.5250 (10) & -2.78 \\
\bottomrule
\end{tabular}
\end{table*}

\subsection{F1-macro and F1-weighted scores} \label{sub:f1scores}

Having investigated the impacts of the few-shot learning and domain knowledge enhancement on prediction accuracy and accuracy-optimal configurations in Sections \ref{sub:accuracy_zero_few} and Section \ref{sub:accuracy_with_witho_domain}, we now extend our analysis to evaluate class-balanced performance through corresponding F1-macro and F1-weighted scores. This analysis is important because accuracy alone can be misleading in imbalanced classification settings, where models may achieve high accuracy while performing poorly on minority classes. F1-macro provides equal weight to each mode, while F1-weighted accounts performance by mode prevalence, thus reflecting overall population-level prediction accuracy.

First, we examine F1 scores for the accuracy-optimized few-shot configurations compared to their zero-shot baselines (Table \ref{tab:zeroshot_vs_fewshot}) to assess whether configurations that maximize accuracy also improve class-balanced performance. Second, we compare F1 scores between accuracy-optimized few-shot configurations with domain knowledge enhancement and without it (Table \ref{tab:domain_impact_fewshot}) to determine whether domain knowledge enhancement that improves accuracy also benefits F1 scores. This approach allows us to investigate whether accuracy optimization aligns with or diverges from F1 optimization, revealing potential trade-offs between overall correctness and class-balanced prediction.

\subsubsection{Performance Comparison on F1 scores: Zero-shot vs One/Few-shot}

Following the accuracy-based analysis in Section~\ref{sub:accuracy_zero_few}, we therefore extend our analysis by examining the F1-macro and F1-weighted scores for these accuracy-optimized configurations to assess whether high accuracy translates to balanced performance across all classes. Table \ref{tab:zeroshot_vs_fewshot} presents the comparative results of F1-macro and F1-weighted scores varying by large language models and sample sizes.

Table \ref{tab:zeroshot_vs_fewshot} shows that there exists a gap between accuracy and F1 scores across all zero-shot configurations with F1-macro around 15-20 percentage lower than accuracy and F1-weighted showing similar but less pronounced gaps. This trend indicates that there exists significant class imbalance where models achieve moderate overall accuracy through bias toward majority classes while underperforming on minority classes. Interestingly, what can be seen clearly from the table is that few-shot learning demonstrates a consistent ability to narrow this accuracy-F1 gap, which demonstrates that in-context examples have the potential to improve class-balanced performance. For example, GPT-4o-mini at a sample size of 100 achieves 124\% F1-macro improvement and 72\% F1-weighted improvement compared to just 28.6\% accuracy improvement, thereby reducing the relative accuracy-F1-macro gap from 49\% to 11\%. This disproportionate F1 enhancement reveals that few-shot learning's primary value lies in teaching models about minority class characteristics rather than simply improving overall correctness. Notably, GPT-4o-mini at a sample size of 200 demonstrates this mechanism more clearly. The prediction accuracy remains unchanged at 0.4750 under both zero-shot and few-shot setting, yet F1-macro increases by 10\% and F1-weighted by 19.8\%. This confirms that few-shot learning can improve class balance independently of overall accuracy. The larger F1-weighted improvement in this case suggests that few-shot learning benefits both minority and majority class predictions when accuracy is already optimized.

However, advanced models exhibit fundamentally different behavior. For instance, o4-mini maintains relatively small gaps across all metrics at the sample size of 1000 (22\% accuracy-F1-macro gap, 6\% accuracy-F1-weighted gap). While few-shot narrows the F1-macro gap, zero-shot achieves superior prediction accuracy and F1-weighted scores. This suggests that advanced models has the ability to pose inherent understanding of class distributions, with zero-shot already achieving strong performance on larger classes (reflected in high F1-weighted) while few-shot provides marginal additional benefits for minority classes (reflected in slightly better F1-macro). 

In summary, the results indicate that few-shot learning acts as an effective mechanism for mitigating class imbalance across both minority and majority classes in traditional models but becomes less necessary for more advanced architectures that have strong pre-trained knowledge of balanced prediction.

\begin{table*}[htbp]
\centering
\caption{Comparison of zero-shot versus best few-shot performance without domain knowledge enhancement}
\label{tab:zeroshot_vs_fewshot}
\begin{tabular}{lccccccc}
\toprule
\multirow{2}{*}{\textbf{Model}} & \multirow{2}{*}{\textbf{Sample Size}} & \multicolumn{3}{c}{\textbf{Zero-Shot}} & \multicolumn{3}{c}{\textbf{Best Few-Shot}} \\
\cmidrule(lr){3-5} \cmidrule(lr){6-8}
 & & \textbf{Accuracy} & \textbf{F1-macro} & \textbf{F1-weighted} & \textbf{Accuracy} & \textbf{F1-macro} & \textbf{F1-weighted} \\
\midrule
\multirow{4}{*}{GPT-4o} 
 & 100 & 0.4000 & 0.2020 & 0.2303 & 0.4000 & 0.2020 & 0.2303 \\
 & 200 & 0.4750 & 0.2354 & 0.3096 & \textbf{0.5250} & \textbf{0.3746} & \textbf{0.3991} \\
 & 500 & 0.4400 & 0.2172 & 0.2828 & 0.4400 & \textbf{0.2194} & \textbf{0.3109} \\
 & 1000 & 0.4500 & 0.2825 & 0.2991 & \textbf{0.4700} & \textbf{0.3618} & \textbf{0.3506} \\
\midrule
\multirow{4}{*}{GPT-4o-mini} 
 & 100 & 0.3500 & 0.1786 & 0.2071 & \textbf{0.4500} & \textbf{0.4007} & \textbf{0.3561} \\
 & 200 & 0.4750 & 0.2354 & 0.3096 & 0.4750 & \textbf{0.2589} & \textbf{0.3708} \\
 & 500 & 0.4000 & 0.1849 & 0.2555 & \textbf{0.4300} & \textbf{0.2216} & \textbf{0.3480} \\
 & 1000 & 0.4350 & 0.2724 & 0.2890 & \textbf{0.4550} & \textbf{0.3399} & \textbf{0.3220} \\
\midrule
\multirow{4}{*}{o3-mini} 
 & 100 & 0.3500 & 0.1911 & 0.2107 & \textbf{0.5000} & \textbf{0.4483} & \textbf{0.4333} \\
 & 200 & 0.4500 & 0.2250 & 0.3000 & \textbf{0.4750} & \textbf{0.2708} & \textbf{0.3427} \\
 & 500 & 0.4400 & 0.2109 & 0.2974 & \textbf{0.4800} & \textbf{0.2988} & \textbf{0.3626} \\
 & 1000 & 0.4500 & 0.3228 & 0.3029 & \textbf{0.4900} & \textbf{0.4569} & \textbf{0.3920} \\
\midrule
\multirow{4}{*}{o4-mini} 
 & 100 & 0.4500 & 0.2887 & 0.4058 & \textbf{0.5000} & \textbf{0.2952} & \textbf{0.4552} \\
 & 200 & 0.5000 & 0.3308 & 0.4745 & \textbf{0.5500} & \textbf{0.4579} & \textbf{0.5339} \\
 & 500 & 0.5600 & 0.4030 & 0.5261 & \textbf{0.5800} & \textbf{0.4743} & \textbf{0.5341} \\
 & 1000 & \textbf{0.5750} & 0.4479 & \textbf{0.5416} & 0.5400 & \textbf{0.4647} & 0.4913 \\
\bottomrule
\end{tabular}
\end{table*}

\subsubsection{Performance Comparison on F1 scores: LLMs with domain knowledge enhancement versus without it}

Following the accuracy-based analysis in Section \ref{sub:accuracy_with_witho_domain}, we therefore extend our analysis by examining F1-macro and F1-weighted scores for these accuracy-optimized configurations to assess whether accuracy improvement translates to balanced performance across all classes. Table \ref{tab:domain_impact_fewshot} presents the comparative results of F1-macro and F1-weighted scores that vary by large language models and sample sizes.

Table \ref{tab:domain_impact_fewshot} demonstrates that domain knowledge enhancement exhibits model-dependent effectiveness, with particularly instructive contrasts between general purpose models (i.e. GPT-4o) and advanced reasoning models (i.e. O4-mini). As can be seen from the table, GPT-4o exhibits the most consistent benefits from domain enhancement, achieving improvements across most configurations in both F1-macro and F1-weighted. For example, at the sample size of 100, domain-knowledge enhancement increases F1-macro by 21.0\% (0.2020 to 0.2444) and F1-weighted by 41.9\% (0.2303 to 0.3267). Notably, these improvements emerge from the interaction between few-shot learning and domain knowledge, where the combination achieves performance gains that neither technique alone could accomplish. This synergistic effect suggests that few-shot examples provide task-specific patterns while domain knowledge supplies contextual understanding, and their integration enables more robust prediction than either approach independently.

However, an interesting and contrasting finding emerges for the advanced reasoning model o4-mini, where domain knowledge enhancement proves largely ineffective or even detrimental, particularly at larger sample sizes. One reason behind this is that advanced reasoning models with sophisticated pre-trained knowledge already possess adequate task understanding while additional domain information may introduce conflicting signals or making some noises to the inherent reasoning pathways of the model.

\begin{table*}[htbp]
\centering
\caption{Impact of domain knowledge enhancement on best few-shot performance}
\label{tab:domain_impact_fewshot}
\begin{tabular}{lccccccc}
\toprule
\multirow{3}{*}{\textbf{Model}} & \multirow{3}{*}{\textbf{Sample Size}} & \multicolumn{3}{c}{\textbf{Without Domain}} & \multicolumn{3}{c}{\textbf{With Domain}} \\
\cmidrule(lr){3-5} \cmidrule(lr){6-8}
 & & \textbf{Accuracy} & \textbf{F1-macro} & \textbf{F1-weighted} & \textbf{Accuracy} & \textbf{F1-macro} & \textbf{F1-weighted} \\
 & & \textbf{(Examples)}  & & &\textbf{(Examples)} &&\\
\midrule
\multirow{4}{*}{GPT-4o} 
 & 100 & 0.4000 (1) & 0.2020 & 0.2303 & \textbf{0.4500 (10)} & \textbf{0.2444} & \textbf{0.3267} \\
 & 200 & 0.5250 (1) & 0.3746 & 0.3991 & \textbf{0.5500 (5)} & \textbf{0.3803} & \textbf{0.4423} \\
 & 500 & 0.4400 (1) & 0.2194 & 0.3109 & \textbf{0.4500 (2)} & \textbf{0.2317} & \textbf{0.3437} \\
 & 1000 & 0.4700 (2) & 0.3618 & 0.3506 & \textbf{0.4850 (3)} & 0.2469 & \textbf{0.3909} \\
\midrule
\multirow{4}{*}{GPT-4o-mini} 
 & 100 & \textbf{0.4500 (2/3)} & \textbf{0.4007} & \textbf{0.3561} & 0.4000 (1) & 0.2203 & 0.2365 \\
 & 200 & 0.4750 (2) & \textbf{0.2589} & \textbf{0.3708} & 0.4750 (5) & 0.2175 & 0.3060 \\
 & 500 & 0.4300 (5) & 0.2216 & 0.3480 & \textbf{0.4900 (5)} & \textbf{0.2586} & \textbf{0.4075} \\
 & 1000 & 0.4550 (2/3/5) & \textbf{0.3399} & \textbf{0.3220} & \textbf{0.4600 (1)} & 0.2342 & 0.3208 \\
\midrule
\multirow{4}{*}{o3-mini} 
 & 100 & \textbf{0.5000 (5)} & \textbf{0.4483} & \textbf{0.4333} & 0.4000 (1) & 0.2203 & 0.2365 \\
 & 200 & 0.4750 (1/3/5/10) & 0.2708 & 0.3427 & \textbf{0.5000 (3)} & \textbf{0.3613} & 0.3518 \\
 & 500 & 0.4800 (2) & 0.2988 & 0.3626 & 0.4800 (3) & \textbf{0.3214} & 0.3448 \\
 & 1000 & \textbf{0.4900 (5)} & \textbf{0.4569} & \textbf{0.3920} & 0.4750 (1) & 0.2537 & 0.3178 \\
\midrule
\multirow{4}{*}{o4-mini} 
 & 100 & \textbf{0.5000 (5)} & 0.2952 & \textbf{0.4552} & 0.4500 (1) & \textbf{0.3278} & 0.4445 \\
 & 200 & 0.5500 (3) & 0.4579 & 0.5339 & \textbf{0.5750 (3)} & \textbf{0.4742} & 0.5168 \\
 & 500 & \textbf{0.5800 (5)} & \textbf{0.4743} & \textbf{0.5341} & 0.5500 (2) & 0.3931 & 0.4934 \\
 & 1000 & \textbf{0.5400 (1)} & \textbf{0.4647} & \textbf{0.4913} & 0.5250 (10) & 0.3414 & 0.4384 \\
\bottomrule
\end{tabular}
\end{table*}

\section{Discussion}

\subsection{Feature engineering for natural language transformation}

This study proposed a feature importance-guided transformation approach that reduces more than 80 variables to fewer than 15 key predictors before linguistic transformation. This dimensionality reduction addresses the practical challenge of generating coherent natural language descriptions from high-dimensional datasets. By focusing narratives on the most predictive variables, this approach constructs interpretable descriptions that capture important behavioral relationships while remaining comprehensible to language models. This approach may extend beyond travel mode prediction to benefit other domains where high-dimensional structured data must be processed through natural language interfaces. The key insight is that feature importance analysis, serving as a critical preprocessing step, can help to narrative descriptions focus on genuinely predictive relationships rather than overwhelming models with peripheral information.

\subsection{Prompting strategies}

The experimental results demonstrate that the effectiveness of prompting strategies in travel mode prediction tasks is tied to model architecture and data availability. Few-shot learning paradigms generally improve the prediction performance but this predictive performance varies by model architecture. The o3-mini model enjoys consistent improvements with few-shot learning. In contrast, o4-mini model often performs better under zero-shot learning compared to few-shot learning. This performance difference suggests that models with advanced reasoning capabilities can extract patterns without requiring additional demonstrations and that few-shot examples may introduce interference rather than guidance. 

Beyond improvements in prediction accuracy, few-shot learning paradigms provide particular value for addressing class imbalance inherent in travel survey data. Under few-shot configurations, models show substantially reduced gaps between accuracy and F1-macro scores, which indicates that demonstration examples help models attend to minority classes rather than simply improving overall prediction accuracy.

Domain knowledge enhancement also follows a complex pattern. GPT-4o benefits consistently (2.27\% to 12.50\% improvements), while o4-mini and o3-mini often perform worse with domain knowledge enhancement, particularly at larger sample sizes. This suggests that more capable models may already encode transportation-relevant relationships through pre-training, and explicit domain prompting creates interference with existing representations. These findings indicate that it should be adaptive and model-aware to conduct domain knowledge integration. General-purpose language models may benefit from domain-knowledge enhancement, while advanced reasoning models may perform optimally with minimal prompting configurations.

\section{Conclusions}

This study proposes TransMode-LLM, a novel framework that leverages the contextual reasoning capabilities of large language models to predict travel mode choice from natural language descriptions. The methodology begins with travel survey data and conducts feature importance analysis to identify the most impacting factors, followed by transforming selected features into descriptive textual inputs. We systematically explore (zero-shot and one/few-shot) and propose a domain-enhanced prompting strategy incorporating standardized mode definitions and a three-step structured decision process. (zero-shot and one/few-shot) and propose a domain-enhanced prompting strategy incorporating standardized mode definitions and a three-step structured decision process. Extensive experiments are conducted on the U.S. National Household Travel Survey (NHTS) across multiple architectures (GPT-4o, GPT-4o-mini, o3-mini, o4-mini) and varying sample sizes.

The experimental results yield three principal findings corresponding to our systematic comparisons. First, the proposed LLM-based approach has the potential to yield competitive performance against established baseline classifiers models (LogitBoost and Gradient Boosting). This finding demonstrates that natural language representations can effectively encode travel behavior information and suggests a paradigm shift from training-based to prompt-based prediction in transportation applications. Second, one/few-shot learning paradigms contribute effectively to improving predictive performance. Typically with 1-5 examples, model achieve accuracy improvements ranging from 3.6\% to 42.9\% over zero-shot learning baselines. Third, general-purpose models (GPT-4o, GPT-4o-mini) benefit from domain knowledge enhancement prompting while for reasoning-oriented models (o3-mini, o4-mini), domain knowledge enhancement does not universally improve performance. 

Given the inherent class imbalance in travel survey data, we build a comprehensive evaluation framework including F1-macro and F1-weighted scores alongside accuracy. This analysis reveals that few-shot learning paradigm's primary value lies in addressing class imbalance rather than simply improving overall accuracy, narrowing the accuracy-F1-macro gap by up to 71\%. 

In summary, these findings carry both theoretical and practical implications. Theoretically, this study bridges traditional quantitative modeling approaches with emerging LLM-based techniques for travel mode prediction, demonstrating that the contextual intelligence of language models can advance travel mode prediction beyond conventional limitations. Practically, the findings highlight the dual potential of large language models in transportation research. LLMs serve not only as powerful predictive tools but also as interpretable frameworks that can provide insights into the multifaceted nature of travel decisions, which also offers significant advantages for transportation planning practitioners and policymakers who require explainable model outputs for decision-making processes.

Several limitations point toward future research directions. First, domain-specific pretraining of language models on transportation field data may yield specialized models better attuned to the nuances of travel behavior modeling. Second, this study did not examine performance variations across demographic groups. Future research should investigate prediction accuracy across age, income, and household composition groups. Third, the current dataset used in this study is from the US, a car-oriented country. To enhance the generalizability, future research should explore data from other countries, particular in in countries oriented to public transport.

\bibliographystyle{unsrt}
\bibliography{references}

\end{document}